\documentclass[%
reprint,
superscriptaddress,
showpacs,preprintnumbers,
 amsmath,amssymb,
aps,
prl,
floatfix%,
]{revtex4-1}

\usepackage{float}
\usepackage{graphicx}% Include figure files
\usepackage{dcolumn}% Align table columns on decimal point
\usepackage{bm}% bold math
\usepackage[multidot]{grffile}
\usepackage[colorlinks=true,allcolors=blue]{hyperref}% add hypertext capabilities
\usepackage{layouts}
\usepackage[utf8]{inputenc}
\usepackage{xcolor}

\begin{document}

%\preprint{APS/123-QED}

\title{Limitations of the asymptotic approach to dynamics}

\author{Julian Newman}
\email{j.newman1@lancaster.ac.uk}
\affiliation{Department of Physics, Lancaster University, Lancaster LA1 4YB, United Kingdom}

\author{Maxime Lucas}
\email{m.lucas@lancaster.ac.uk}
\affiliation{Department of Physics, Lancaster University, Lancaster LA1 4YB, United Kingdom}
\affiliation{Dipartimento di Fisica e Astronomia, Universit{\`a} di Firenze, INFN and CSDC, Via Sansone 1, 50019 Sesto Fiorentino, Firenze, Italy}

\author{Aneta Stefanovska}
\email{aneta@lancaster.ac.uk}
\affiliation{Department of Physics, Lancaster University, Lancaster LA1 4YB, United Kingdom}

\date{\today}% It is always \today, today,

\begin{abstract}
	Standard dynamical systems theory is centred around the coordinate-invariant asymptotic-time properties of autonomous systems. We identify three limitations of this approach. Firstly, we discuss how the traditional approach cannot take into account the time-varying nature of dynamics of open systems. Secondly, we show that models with explicit dependence on time exhibit stark dynamic phenomena, even when they cannot be defined for infinite time. We see a bifurcation occurring in nonautonomous finite-time systems that cannot be identified by classical methods for infinite-time autonomous systems. Thirdly, even when a time-varying model can be extended to infinite time, the classical infinite-time approach is likely to miss dynamical phenomena that are more readily understood within the framework of finite-time dynamics. We conclude the potentially crucial importance of a nonautonomous finite-time approach to real-world, open systems.
\end{abstract}

\pacs{05.45.Xt, 05.65.+b, 89.75.Fb}% PACS, the Physics and Astronomy
\maketitle

Dynamics, as introduced by Newton to describe celestial motion, and later extended by Lyapunov~\cite{Lyapunov1892} and Poincar\'{e}~\cite{Poincare1881} to include stability analysis, has been fruitfully applied to various fields such as control and communication, and many inverse approach methods have been developed to extract features of dynamical systems from measured data. Two popular branches of dynamical systems theory are deterministic chaotic dynamics~\cite{strogatz1994} and synchronization of interacting dynamical processes~\cite{Pikovsky2001}. Still, all of this has mostly, to date, been based on two main assumptions, that the dynamics of the systems is time-homogeneous, and that the physical behaviour exhibited can be described by coordinate-invariant time-asymptotic dynamics.

However, many real-world systems are \emph{open} and thus too prone to time-variable influences to be reasonably modelled by a time-independent evolution law~\cite{kloeden2011,kloeden2013}. Examples where this can be the case include the firing of neurons~\cite{cressman2009}, the cardiovascular system~\cite{Stefanovska2000}, the climate~\cite{desaedeleer2013, Perryman2014,Moon2017b}, metabolic oscillations~\cite{gustavsson2012,lancaster2016}, circadian rhythms~\cite{flourakis2015}, and quantum mechanics \cite{serkin2007}.

Finite-time nonautonomous dynamics has recently been gaining attention~\cite{Haller2000,Lekien2007,Berger2008,Rasmussen2010,Karrasch2013,Giesl2018}. In~\cite{ushio2018}, which analysed multi-species population dynamics on the basis of real data, the findings relied on a quantitative measure of time-evolving dynamical stability which would have not been possible within the framework of time-asymptotic dynamics. Many diverse contexts involving fluid flows have been investigated analytically, numerically and from data ~\cite{Haller2000b,Shadden2006,Lehahn2007,TewKai2009,Toger2012,moorcroft2013,Ivic2017,Wang2017,Ramos2018}, in order to study \emph{coherent structures} within the body of fluid, such as the Red Spot on Jupiter~\cite{Haller2015}. These structures are typically identified in terms of finite-time Lyapunov exponents (FTLE), and exist completely independently of whether temporal variations follow an infinitely extendible pattern over time -- which, typically, they do not.

In this Letter, we demonstrate the limitations of the above-mentioned two assumptions by uncovering a dynamical phenomenon that cannot be described by the standard approach based on these assumption. Specifically, we show that for very general slowly varying one-dimensional phase-oscillator systems, sufficiently broad variation inherently induces stability in the system.

Consider a differential equation
\begin{equation} \label{ode} \dot{\theta}(t) \ = \ F(\theta(t),\tfrac{t}{T}) \end{equation}
on $\mathbb{S}^1=\mathbb{R}/(2\pi\mathbb{Z})$, defined on time-interval $[0,T]$, where $F \colon \mathbb{S}^1 \times [0,1] \to \mathbb{R}$ is a smooth function. Eq.~\eqref{ode} is a time-varying differential equation, where \emph{$F$ specifies the shape} of the variation, while \emph{$T$ specifies the slowness} at which this shape of variation is realised. Eq.~\eqref{ode} can describe various physical situations, due to existence of phase reduction methods for slowly varying systems~\cite{kurebayashi2013,kurebayashi2015,park2016}. At any time $t \in [0,T]$, an \emph{instantaneous stable equilibrium of \eqref{ode}} means a point $y_{\frac{t}{T}} \in \mathbb{S}^1$ such that $F(y_{\frac{t}{T}},\frac{t}{T})=0$ and $\frac{\partial F}{\partial \theta}(y_{\frac{t}{T}},\frac{t}{T})<0$.

We consider two cases:
\begin{itemize}
\item Case~I: either $F(\theta,\tau)>0$ for all $\theta$ and $\tau$, or $F(\theta,\tau)<0$ for all $\theta$ and $\tau$.
\item Case~II: there exist times $t \in [0,T]$ at which \eqref{ode} has a unique instantaneous stable equilibrium.
\end{itemize}
For generic $F$, if the function $F(\,\cdot\,,0)$ or $F(\,\cdot\,,1)$ has no zeros then the system is either in Case~I or Case~II. Our results, generalising results of \cite{Jensen2002,Lucas2018}, can be summarised as follows. \emph{Assuming slow variation of $F$: in Case~I, \eqref{ode} exhibits neutrally stable dynamics; in Case~II, typically, \eqref{ode} exhibits global-scale stable dynamics.} The neutral stability in Case~I means that there is no significant attractivity or repulsivity of the solutions. The global-scale stability in Case~II means that all solutions starting outside some very small ``repulsive'' arc will cluster together over time into a very small arc. The transition from Case~I to Case~II resembles a classical saddle-node bifurcation.

The result can be explained as follows: While the instantaneous vector field $F(\,\cdot\,,\frac{t}{T})$ has no zeros, slow variation implies that trajectories move approximately periodically round the circle, exhibiting no significant mutual synchronisation or repulsion. While $F(\,\cdot\,,\frac{t}{T})$ has an instantaneous stable equilibrium, slow variation implies that trajectories have time to reach it and follow its slow motion, clustering into an increasingly tight cluster around it.

Throughout the following, an \emph{arc} $J \subset \mathbb{S}^1$ is assumed to be a closed connected proper subset of $\mathbb{S}^1$ with non-empty interior. Given an arc $J_0 \subset \mathbb{S}^1$ of initial conditions $\theta(0)$, we write $J_t$ for the arc of subsequent positions $\theta(t)$ at time $t$. The result for Case~I can be mathematically formalised~\footnote{The proof, based on an application of Grönwall's inequality~\cite{Dragomir2003}, is given in the Supplementary Material.} as follows.

\textbf{Proposition 1.} \emph{Fix any $F$ within Case~I. There exists a constant $c_F \geq 1$ independent of $T$, such that for every arc $J_0$, for all $t \in [0,T]$,}
\[ \frac{1}{c_F} \ \leq\ \frac{\mathrm{length}(J_t)}{\mathrm{length}(J_0)} \ \leq\ c_F. \]
An explicit formula for $c_F$ is given in the Supplementary Material \footnote{One can also obtain Proposition~1 (without the formula for $c_F$) as a consequence of \cite[Theorem~2]{Guckenheimer2001}}. Proposition~1 implies in particular that the FTLE associated to all trajectories over $[0,T]$ are bounded in absolute value by $\frac{1}{T}\log c_F$, and thus these FTLE tend to $0$ as $T \to \infty$.

In Case~II, if there is only one time-interval during which an instantaneous stable equilibrium exists, the stability can be mathematically formalised and quantified \footnote{The proof is similar in nature to the proof of \cite[Proposition~4]{Guckenheimer2001}; it is based on adiabatic reasoning, such that in principle one could construct an explicit formula defining ``sufficiently large'' $T$. The full proof, while not requiring any deep insight, is nonetheless lengthy and technical, and will be published in a future work.} as follows.

\textbf{Proposition 2.} \emph{Fix $F$ such that there exist $0 \leq \tau_1 < \tau_2 \leq 1$ satisfying}:\emph{
\begin{itemize}
\item for all $\tau \in [0,\tau_1) \cup (\tau_2,1]$ and $\theta \in \mathbb{S}^1$, we have $F(\theta,\tau) \neq 0$;
\item there is a continuous map $\tau \mapsto y_\tau$ from $[\tau_1,\tau_2]$ to $\mathbb{S}^1$ such that for each $\tau \in (\tau_1,\tau_2)$, we have $F(y_\tau,\tau)=0$ and $\frac{\partial F}{\partial \theta}(y_\tau,\tau)<0$;
\item there exists $0<\delta \leq \tau_2-\tau_1$ and a continuous map $\tau \mapsto z_\tau$ from $[\tau_1,\tau_1+\delta]$ to $\mathbb{S}^1$ such that for each $\tau \in (\tau_1,\tau_1+\delta]$, we have $F(z_\tau,\tau)=0$, $\frac{\partial F}{\partial \theta}(z_\tau,\tau)>0$ and $F(\theta,\tau) \neq 0$ for all $\theta \in \mathbb{S}^1 \setminus \{y_\tau,z_\tau\}$.
\end{itemize}
Let
\begin{equation} \label{le} \Lambda \ := \ \int_{\tau_1}^{\tau_2} \partial_1F(y_\tau,\tau) \, d\tau \ < \ 0. \end{equation}
Fix any $\varepsilon>0$. Then, provided $T$ is sufficiently large, there exists an arc $P$ with $\mathrm{length}(P)<\varepsilon$ such that for every arc $J_0$ not intersecting $P$,}
\[ \left| \frac{1}{T} \log\left( \frac{\mathrm{length}(J_T)}{\mathrm{length}(J_0)} \right) \ - \ \Lambda \right| \, < \, \varepsilon. \]
The quantity $\Lambda$ defined in \eqref{le} is an approximation of the FTLE over $[0,T]$ associated to all trajectories except those starting in a small arc $P$.

\begin{figure*}[hbt!]
	\centering	\includegraphics[width=\linewidth]{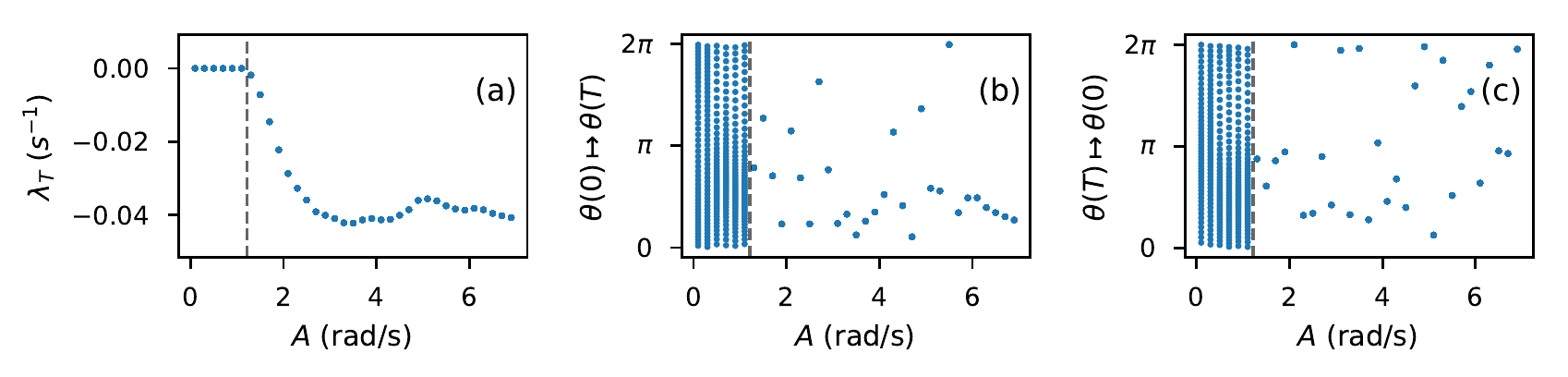}
	\caption{Stabilisation of \eqref{example} as $A$ is increased. Other parameters are $a=\frac{1}{3}\;\mathrm{rad}/\mathrm{s}$ and $k=1\;\mathrm{rad}/\mathrm{s}$; the cut-off frequency for the lowpass filter in the construction of $g(t)$ is $1/(2\pi \times 10^3)\;$Hz. In (a) and (b), for each $A$-value, results for the evolution $\theta(t)$ of 50 equally spaced initial conditions $\theta(0)=\frac{2\pi i}{50}$, $i=0,\ldots,49$, are shown: (a) shows the FTLE $\lambda_T$, as defined by \eqref{ftle}, for these trajectories; (b) shows the positions $\theta(T)$ of these trajectories at time $T$. In (c), for each $A$-value, the positions of $\theta(0)$ for the 50 trajectories ending at the points $\theta(T)=\frac{2\pi i}{50}$, $i=0,\ldots,49$, are shown. The value $A^\ast$ as defined in \eqref{star1} is marked in dashed black.}
	\label{mainfig}
\end{figure*}

If there is more than one time-interval during which an instantaneous stable equilibrium exists, then generally this will just further reinforce the mutual synchrony of trajectories. However, it is also theoretically possible that the cluster of trajectories formed over one or more of these time-intervals will happen to land in the small repulsive arc associated to the next of these time-intervals, causing the cluster to be re-dispersed. Generally (apart from some degenerate classes of examples), this behaviour will be very rare and will require extremely fine tuning of parameters.

We now consider two simple examples, both of the particular form considered in \cite{Jensen2002,Lucas2018} where $\theta$ models the phase difference in a unidirectionally coupled pair of oscillators, with the frequency of the driving oscillator slowly varied. Firstly, we consider on the time-interval $[0,T]$ with $T=2\pi \times 10^5\;$s the differential equation
\begin{equation} \label{example} \dot{\theta}(t) \ = \ -a\sin(\theta(t)) + k + A g(t) \end{equation}
for fixed $k>a>0$, where $g \colon [0,T] \to \mathbb{R}$ is the output of passing a sample realisation of a Brownian bridge through a lowpass filter. As in \cite{Lucas2018}, the value $A>0$, which we leave free, quantifies the ``breadth'' of variation.  Since $g(t)$ is constructed from a Brownian bridge defined on a finite time-interval, there is no meaningfully natural way to extend the definition of $g(t)$ to infinite time. Thus, \textit{asymptotic-dynamics concepts and methods are inapplicable}.

Provided the range of $g(t)$ includes negative values, we define the \emph{critical $A$-value} $A^\ast$ to be largest $A$-value such that $-a\sin(\theta)+k+Ag(t) \geq 0$ for all $\theta$ and $t$, namely
\begin{equation} \label{star1} A^\ast \ = \ \frac{a-k}{\min_{0 \leq t \leq T} g(t)}. \end{equation}
For $A<A^\ast$, the system is in Case~I and Proposition~1 applies: we should expect no significant mutual synchronisation of trajectories; the FTLE $\lambda_T$ associated to each trajectory $(\theta(t))_{0 \leq t \leq T}$, defined by
\begin{equation} \label{ftle} \lambda_T \ =  \ \frac{1}{T} \int_0^T -a \cos(\theta(t)) \, dt, \end{equation}
should be approximately zero. For $A>A^\ast$, the system is in Case~II. We should expect the trajectories of virtually all initial conditions to be clustered together around one point at time $T$, being repelled away from the small vicinity of a repulsive point; the FTLE as defined by \eqref{ftle} should be negative for all trajectories starting outside the small repulsive region. 

Numerical results are shown in Fig.~\ref{mainfig}; a detailed description of these numerics, as well as some further numerics, are given in the Supplementary Material. For each $A < A^\ast$, we see trajectories spread throughout the circle (plot~(b)), all with zero FTLE (plot~(a)). For each $A > A^\ast$, we see trajectories clustered around one point at time $T$ (plot~(b)), all with a shared negative FTLE (plot~(a)), and repelled away from the vicinity of one point (plot~(c)). This picture strongly resembles the classical saddle-node bifurcation for autonomous dynamical systems, but cannot be obtained by traditional asymptotic dynamics analysis.

So then, physically, increasing the breadth of variation $A$ of the nonautonomous influence makes the system stable, when it was merely neutrally stable. This stabilisation phenomenon, which did not depend on a specific form of $g$, highlights a surprising and potentially very important connection between time-variability and stability. Since all open systems are subject to the effects of time-variable influences, temporal variation may play a key role in the mechanisms by which some systems maintain stable functioning~\cite{Stefanovska2000,suprunenko2013}.

%fig1

As a second example, we consider the differential equation
\begin{equation} \label{sinus} \dot{\theta}(t) \ = \ -a\sin(\theta(t)) + k + A \cos(\omega t) \end{equation}
with $a,\omega>0$ and $k,A \geq 0$. This is essentially the same as used for the numerics in \cite{Lucas2018}. It has also been studied in \cite{Guckenheimer2001,Ilyashenko2011,Gandhi2015,Buchstaber2017} and references therein: as well as the phase difference of unidirectionally coupled oscillators, Eq.~\eqref{sinus} also models a resistively shunted Josephson junction driven by biased alternating current.

Slow variation here means that $A\omega$ is small. If we fix $k>a$, and consider Eq.~\eqref{sinus} over a time-interval $[0,T]$ with $T>\frac{\pi}{\omega}$, then the critical $A$-value is $A^\ast = k-a$: for $A<A^\ast$ we have neutrally stable dyamics, and for $A>A^\ast$ we have stable dynamics, all exactly as with~\eqref{example}. As in \cite{Jensen2002} and Eq.~\eqref{le} above, one can derive an approximation $\tilde{\Lambda}$ for FTLE as defined by \eqref{ftle} over time-intervals of integer period lengths $T=\frac{2\pi n}{\omega}$, by adiabatically following the slowly moving attracting point when it exists, i.e.
\begin{equation} \label{l0} \tilde{\Lambda} = \frac{1}{\pi} \int_{\{0 \leq s \leq \pi \, : \, |k+A\cos(s)| < a \}} \!\!\! -a\cos(y_\tau) \, d\tau \end{equation}
where $y_\tau=\mathrm{arc}\sin\left(\frac{k+A\cos(\tau)}{a}\right)$. For $A<k-a$, we have $\tilde{\Lambda}=0$, and for $A>k-a$, we have $\tilde{\Lambda}<0$. Extensive numerics in the Supplementary Material show the same bifurcation scenario as for \eqref{example}, with FTLE being approximated well by $\tilde{\Lambda}$ (as also in Fig.~\ref{fig:fig2} below).

\begin{figure}[hbt]
	\centering
	\includegraphics[width=\linewidth]{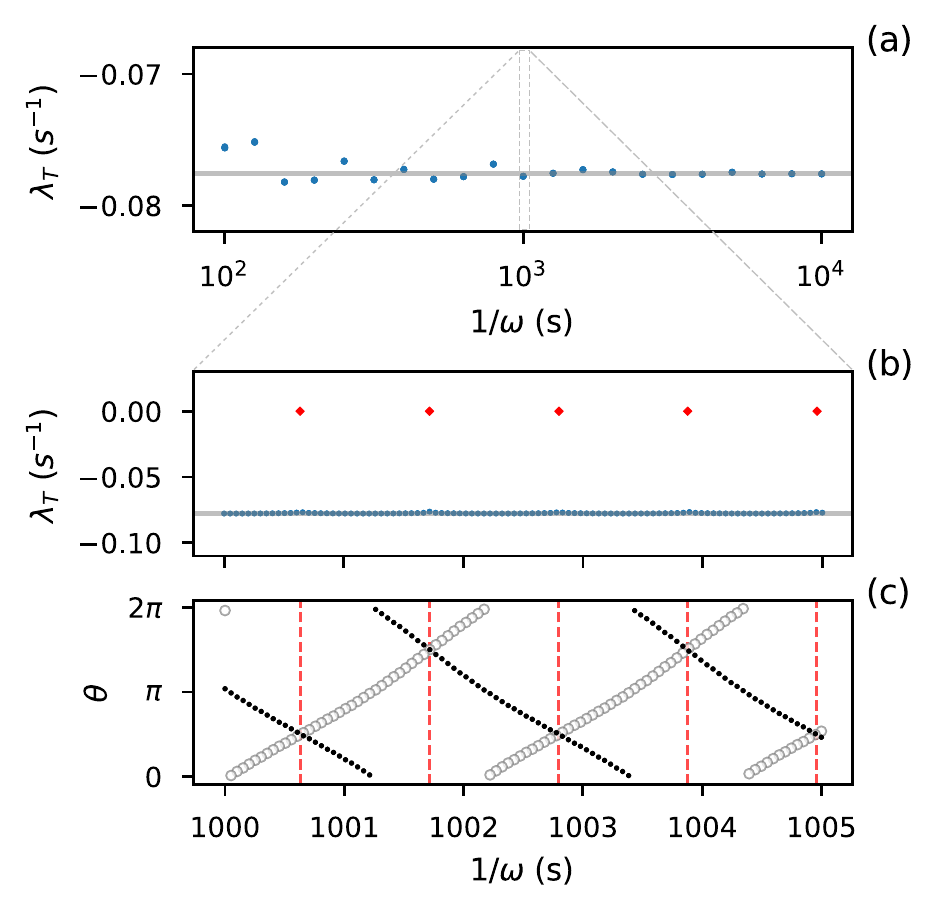}
	
	\caption{Discrepancy between finite-time analysis and asymptotic analysis, in system~\eqref{sinus}, as $\omega$ is varied. Other parameters are $a=\frac{1}{3}\;\mathrm{rad}/\mathrm{s}$ and $A=k=1\;\mathrm{rad}/\mathrm{s}$. (a) For each $\omega$-value, the FTLE $\lambda_T$ as defined by \eqref{ftle} are shown for the trajectories of 50 equally spaced initial conditions $\theta(0)=\frac{2\pi i}{50}$, $i=0,\ldots,49$, with $T=\frac{200\pi}{\omega}$ (i.e.\ $100$ periods). The predicted approximate FTLE $\tilde{\Lambda}<0$ defined in \eqref{l0} is marked in grey. The FTLE are indeed approximately equal to $\tilde{\Lambda}$ over the whole $\omega$-range, especially for smaller $\omega$. (b) Zoomed-in version of (a); the red points indicate the location of the small intervals of $\omega$-values for which all initial conditions have zero asymptotic Lyapunov exponent. (c) Forward (hollow circle) and backward (solid circle) evolution of 50 equally spaced points $\frac{2\pi i}{50}$, $i=0,\ldots,49$, over the time-interval $[0,\frac{2\pi}{\omega}]$. In both forward and backward time, the 50 trajectories cluster around one point; the values of $\omega$ where the two curves of clustered points cross correspond to where the red points are marked in (b); see the Supplementary Material for further explanation.}
	\label{fig:fig2}
\end{figure}

But unlike in \eqref{example}, the nonautonomous term $A\cos(\omega t)$ in Eq.~\eqref{sinus} happens to be periodic; therefore, it is possible to treat \eqref{sinus} as an infinite-time system and analyse Eq.~\eqref{sinus} within the traditional framework of coordinate-invariant asymptotic dynamics, just as most studies of Eq.~\eqref{sinus} have done. The notions of ``stable dynamics'' and ``neutrally stable dynamics'' can be formalised within the traditional asymptotic framework. From this point of view, one obtains the following basic fact \footnote{This follows from \cite[Theorems~1 and 4]{Ilyashenko2011}. See the Supplementary Material for further details.}.

\textbf{Proposition 3.} \emph{Fix $k>a>0$. For any $A>0$, there are intervals of $\omega$-values arbitrarily close to $0$ for which Eq.~\eqref{sinus} exhibits neutrally stable dynamics, with all trajectories having an asymptotic Lyapunov exponent of exactly zero.}

 So, the asymptotic approach gives that for any $A > 0$, one can find arbitrarily small $\omega$-values for which the system is neutrally stable. This stands in contrast to our above adiabatic approach where, provided $A\omega$ is small, $A>k-a$ implies stable dynamics. This discrepancy between the two approaches is shown in Fig.~\ref{fig:fig2}. The reason for the discrepancy is the re-dispersion effect described further above; as indicated in the Supplementary Material, it requires extreme fine-tuning and is not observed in our numerical simulations. Therefore, the existence of these $\omega$-intervals of zero asymptotic Lyapunov exponent is of far less physical relevance, if any, than the general stabilisation that we have described for when $A$ rises above $k-a$. But the mathematical tools needed to obtain Proposition~3 do not reveal any of this information. This highlights the danger that one seeking help from traditional dynamical systems and bifurcation theory in order to understand real-world, open systems could be misled.

 Thus, in this paper, we have seen that restricting the analysis of dynamics to the traditional framework has the potential to impede progress in diverse fields of scientific inquiry, such as all those mentioned further above. The time-variable and finite-time nature of open systems needs to be incorporated in the development and application of dynamical systems theory. A severe misconception is that standard autonomous dynamical systems theory automatically covers the need to understand nonautonomous dynamics, since the introduction of time into the phase space~\cite{strogatz1994} as a variable $\tau$ with $\dot{\tau}=1$ makes the nonautonomous system autonomous. However, the qualitative behaviour of this autonomised version of a nonautonomous system is trivial from the point of view of standard autonomous theory: all solutions simply move towards $\infty$. Autonomous theory generally focuses on \emph{bounded} objects such as fixed points, periodic orbits, invariant measures and associated Lyapunov exponents, etc.; but none of these exist for a system involving a component $\dot{\tau}=1$~\cite[Remark~2.5]{kloeden2011}. Moreover, the flows of this autonomised version of any two nonautonomous systems on $\mathbb{R}^N$ can be conjugated to each other by a $\tau$-preserving diffeomorphism of the extended space; so all coordinate-invariant dynamics is destroyed by this autonomisation approach.

In conclusion, a nonautonomous finite-time approach to real-world, open systems is potentially of crucial importance. The reality of unignorable time-variability also has implications for inverse problem methodologies; \emph{time-localised} analysis and inference methods~\cite{Clemson2014,Stankovski2012}, which do not treat all temporal variations as noise, will not only reveal more information than their time-independent counterparts but also allow for much more reliable conclusions about systems that may be time-varying.

The authors would like to express their gratitude to  Bostjan Dolenc, Edgar Knobloch, Peter McClintock, Woosok Moon, Arkady Pikovsky,  Antonio Politi, Martin Rasmussen, Janne Ruostekoski, David Sloan, and Yevhen Suprunenko, for interesting and insightful discussions. This study has been supported by an EPSRC Doctoral Prize Fellowship, the DFG grant CRC 701 \emph{Spectral Structures and Topological Methods in Mathematics}, the EPSRC grant EP/M006298/1 \emph{A device to detect and measure the progression of dementia by quantifying the interactions between neuronal and cardiovascular oscillations}, and the European Union's Horizon 2020 research and innovation programme under the Marie Sk\l{}odowska-Curie grant agreement No 642563.

\bibliography{bib11}

\end{document}

%\preprint{APS/123-QED}

\title{Limitations of the asymptotic approach to dynamics: Supplementary Material}

\author{Julian Newman}
\email{j.newman1@lancaster.ac.uk}
\affiliation{Department of Physics, Lancaster University, Lancaster LA1 4YB, United Kingdom}

\author{Maxime Lucas}
\email{m.lucas@lancaster.ac.uk}
\affiliation{Department of Physics, Lancaster University, Lancaster LA1 4YB, United Kingdom}
\affiliation{Dipartimento di Fisica e Astronomia, Universit{\`a} di Firenze, INFN and CSDC, Via Sansone 1, 50019 Sesto Fiorentino, Firenze, Italy}

\author{Aneta Stefanovska}
\email{aneta@lancaster.ac.uk}
\affiliation{Department of Physics, Lancaster University, Lancaster LA1 4YB, United Kingdom}

\date{\today}% It is always \today, today,

\pacs{05.45.Xt, 05.65.+b, 89.75.Fb}% PACS, the Physics and Astronomy
\maketitle

In this Supplementary Material, we will:
\begin{itemize}
\item give a proof of Proposition~1 of the main text, including an explicit formula for the bound $c_F$;
\item explain Proposition~3 and how it is proved;
\item describe carefully how the numerics in Fig.~1 of the main text were obtained;
\item further support the stabilisation phenomenon described in the main text, by carrying out the same numerics on (3) but ending at an earlier time, and seeing the same bifurcation at the corresponding predicted critical $A$-value;
\item describe carefully how the numerics in Fig.~2 of the main text were obtained, and explain why the red points marked in Fig.~2(b) correspond to small intervals where the asymptotic Lyapunov exponent (ALE) is zero;
\item further support the stabilisation phenomenon for the system (6) as described in the main text, by numerics showing the predicted behaviour for varying $A$, varying $k$, growing $t$, and different $\omega$.
\end{itemize}

\section*{Proof of Proposition 1}

Define
\begin{align*}
\mathfrak{e}(x) \ &= \ \sum_{i=0}^\infty \frac{x^i}{(i+3)!} \\
&= \ \frac{e^x-(1+x+\frac{1}{2}x^2)}{x^3} \hspace{3mm} \textrm{if }x \neq 0
\end{align*}
and let
\[ M_F \ = \ \max_{\tau \in [0,1]} \int_{\mathbb{S}^1} \frac{1}{|F(\theta,\tau)|} \, d\theta. \]
Given a continuous function $G \colon \mathbb{S}^1 \times [0,1] \to \mathbb{R}$, we write $\|G\|:=\max_{\theta,\tau} |G(\theta,\tau)|$. We prove Proposition~1 with the bound
\[ \boxed{\  c_F \ = \ \mathrm{exp}\left( M_F^2 \mathfrak{e}(M_F\|\tfrac{\partial F}{\partial \theta}\|)\|\tfrac{\partial^2 F}{\partial \theta^2}\|\|\tfrac{\partial F}{\partial \tau}\| + \tfrac{1}{2}M_F\|\tfrac{\partial^2F}{\partial\theta\partial\tau}\| + M_F\|\tfrac{\partial F}{\partial \theta}\| \right). \; } \]

Assume without loss of generality that $F(\theta,\tau)>0$ for all $\theta$ and $\tau$, and that $F(\theta,\tau)$ is not independent of $\theta$. For each $0 \leq s \leq t \leq T$, write $f_{s,t} \colon \mathbb{S}^1 \to \mathbb{S}^1$ for the map sending $\theta(s)$ to $\theta(t)$ for all solutions $\theta(\cdot)$ of Eq.~(1) in the main text. Let $0=t_0 < t_1 < \ldots < t_N$ be such that for each $i<N$,
\[ t_{i+1}-t_i \ = \ \int_{\mathbb{S}^1} \frac{1}{F(\theta,\frac{t_i}{T})} \, d\theta, \]
with $N$ being the largest possible integer such that $t_N \leq T$. Note that $t_{i+1}-t_i \leq M_F$ for each $i<N$, and $T-t_N < M_F$. For each $i<N$, write $\tilde{f}_{i,s,t} \colon \mathbb{S}^1 \to \mathbb{S}^1$ for the map sending $\psi(s)$ to $\psi(t)$ for all solutions $\psi(\cdot)$ of the differential equation
\begin{equation} \label{av} \dot{\psi}(t) \ = \ F(\psi(t),\tfrac{t_i}{T}). \tag{*} \end{equation}
Take any $i < N$, let $\hat{\theta} \colon [0,T] \to \mathbb{R}$ be a lift of any solution $\theta(\cdot)$ of (1) and let $\hat{\psi} \colon \mathbb{R} \to \mathbb{R}$ be the lift of a solution $\psi(\cdot)$ of \eqref{av} such that $\hat{\psi}(t_i)=\hat{\theta}(t_i)$. Then for all $t \in [t_i,t_{i+1}]$,
\begin{align*}
|\hat{\theta}(t) - \hat{\psi}(t)| \ &\leq \ \int_{t_i}^t |F(\theta(s),\tfrac{s}{T}) - F(\psi(s),\tfrac{t_i}{T})| \, ds \\
&\leq \ \int_{t_i}^t \tfrac{1}{T}\|\tfrac{\partial F}{\partial \tau}\|(s-t_i) \ + \ \|\tfrac{\partial F}{\partial \theta}\| |\hat{\theta}(s) - \hat{\psi}(s)| \, ds
\end{align*}
and so a suitable version of Gr\"{o}nwall's Inequality~{[49,~Corollary~2]} gives that
\begin{align*}
|\hat{\theta}(t) - \hat{\psi}(t)| \ &\leq \ \int_{t_i}^t \tfrac{1}{T}\|\tfrac{\partial F}{\partial \tau}\|(s-t_i) e^{\|\frac{\partial F}{\partial \theta}\|(t-s)} \, ds \\
&= \ \tfrac{1}{T}\|\tfrac{\partial F}{\partial \tau}\| \|\tfrac{\partial F}{\partial \theta}\|^{-2}\left(e^{\|\frac{\partial F}{\partial \theta}\|(t-t_i)} - (1 + \|\tfrac{\partial F}{\partial \theta}\|(t-t_i))\right)
\end{align*}
provided $F(\,\cdot\,,\tau)$ is not a constant function for all $\tau$. Consequently, using the fact that all solutions of \eqref{av} are $(t_{i+1}-t_i)$-periodic, we have that
\begin{align*}
|\log f_{t_i,t_{i+1}}' (\theta(t_i))| \ &= \ |\log f_{t_i,t_{i+1}}' (\theta(t_i)) - \log \tilde{f}_{i,t_i,t_{i+1}}' (\theta(t_i))| \\
&= \ \left| \int_{t_i}^{t_{i+1}} \tfrac{\partial F}{\partial \theta}(\theta(t),\tfrac{t}{T}) - \tfrac{\partial F}{\partial \theta}(\psi(t),\tfrac{t_t}{T}) \, dt \right| \\
&\leq \ \frac{1}{T} \int_{t_i}^{t_{i+1}} \|\tfrac{\partial^2F}{\partial\theta^2}\| \|\tfrac{\partial F}{\partial \tau}\| \|\tfrac{\partial F}{\partial \theta}\|^{-2}\left(e^{\|\frac{\partial F}{\partial \theta}\|(t-t_i)} - (1 + \|\tfrac{\partial F}{\partial \theta}\|(t-t_i))\right) \,+ \, \|\tfrac{\partial^2F}{\partial\theta\partial\tau}\|(t-t_i) \, dt \\
&= \ \frac{1}{T}\left( \|\tfrac{\partial^2F}{\partial\theta^2}\| \|\tfrac{\partial F}{\partial \tau}\| \mathfrak{e}(\|\tfrac{\partial F}{\partial \theta}\|(t_{i+1}-t_i))(t_{i+1}-t_i)^3 + \tfrac{1}{2}\|\tfrac{\partial^2F}{\partial\theta\partial\tau}\|(t_{i+1}-t_i)^2 \right) \\
&\leq \ \frac{t_{i+1}-t_i}{T} \left( \|\tfrac{\partial^2F}{\partial\theta^2}\| \|\tfrac{\partial F}{\partial \tau}\| \mathfrak{e}(M_F\|\tfrac{\partial F}{\partial \theta}\|)M_F^2 + \tfrac{1}{2}\|\tfrac{\partial^2F}{\partial\theta\partial\tau}\|M_F \right).
\end{align*}
This is true for any solution $\theta(\cdot)$ of (1). Combining this with the fact that
\[ |\log f_{t_N,T}'(\theta(t_N))| \ = \ \left|\int_{t_N}^{T} \tfrac{\partial F}{\partial \theta}(\theta(t),\tfrac{t}{T}) \, dt \right| \ \leq \ M_F\|\tfrac{\partial F}{\partial \theta}\|, \]
we obtain that
\[ |\log f_{0,T}'(\theta(0)) | \ \leq \ M_F^2 \mathfrak{e}(M_F\|\tfrac{\partial F}{\partial \theta}\|)\|\tfrac{\partial^2 F}{\partial \theta^2}\|\|\tfrac{\partial F}{\partial \tau}\| + \tfrac{1}{2}M_F\|\tfrac{\partial^2F}{\partial\theta\partial\tau}\| + M_F\|\tfrac{\partial F}{\partial \theta}\| \ = \ \log c_F. \]
Since this is true for every solution $\theta(\cdot)$, it follows that for every arc $J_0 \subset \mathbb{S}^1$,
\[ \frac{1}{c_F} \ \leq\ \frac{\mathrm{length}(J_T)}{\mathrm{length}(J_0)} \ \leq\ c_F. \]

Now for any $t \in (0,T)$, we can define the ``restricted'' function $F[\frac{t}{T}] \colon \mathbb{S}^1 \times [0,1] \to \mathbb{R}$ by
\[ F[\tfrac{t}{T}](\theta,\tau) \ = \ F(\theta,\tfrac{\tau t}{T}). \]
Observe that $c_{F[\frac{t}{T}]} \leq c_F$, and therefore
\[ \frac{1}{c_F} \ \leq\ \frac{\mathrm{length}(J_t)}{\mathrm{length}(J_0)} \ \leq\ c_F \]
for all $t \in [0,T]$.

\section*{Proposition 3}

As a slight generalisation of Eq.~(6), we can consider
\begin{equation} \label{general} \dot{\theta}(t) \ = \ -a\sin(\theta(t)) + k + A p(\omega t) \tag{**} \end{equation}
where $p(\cdot)$ is any smooth $2\pi$-periodic function satisfying $\int_0^{2\pi} p(s) \, ds = 0$.

Once again, an \emph{arc} is a closed connected proper subset of $\mathbb{S}^1$ with non-empty interior. Given an arc $J_0 \subset \mathbb{S}^1$ of initial conditions, we write $J_t$ for the arc of positions of the subsequent trajectories of \eqref{general} at time $t$. By {[43,~Theorems~1 and 4]}, from the point of view of coordinate-invariant asymptotic dynamics, Eq.~\eqref{general} may exhibit neutrally stable dynamics or global-scale stable dynamics, otherwise it must exhibit dynamics lying at the ``boundary'' between these two. More precisely, the three possible scenarios are as follows:
\begin{itemize}
\item \emph{Neutrally Stable Scenario}: There exists $c \geq 1$ such that for every arc $J_0$, for all $t \in [0,\infty)$,
\[ \frac{1}{c} \ \leq\ \frac{\mathrm{length}(J_t)}{\mathrm{length}(J_0)} \ \leq\ c. \]
In this case, the ALE associated to all trajectories is $0$.
\item \emph{Stable Scenario}: There exists $\lambda<0$ and $p \in \mathbb{S}^1$ such that for every arc $J_0$ with $p \notin J_0$,
\[ \frac{1}{t} \log\left( \frac{\mathrm{length}(J_t)}{\mathrm{length}(J_0)} \right) \to \lambda \ \ \textrm{ as } \ t \to \infty. \]
In this case, the ALE associated to every trajectory except the trajectory starting at $p$ is $\lambda$. The trajectory starting at $p$ is an unstable $\frac{2\pi}{\omega}$-periodic trajectory, and all other trajectories are attracted to a stable $\frac{2\pi}{\omega}$-periodic trajectory.
\item \emph{Boundary Scenario}: There is a $\frac{2\pi}{\omega}$-periodic solution which asymptotically attracts all trajectories from one direction but is unstable due to being locally repulsive in the other direction. In this case, the ALE associated to every trajectory is $0$.
\end{itemize}
The first two scenarios are asymptotic-dynamical analogues of the dynamics described in Propositions~1 and 2 respectively. Proposition~3 asserts that if $k>a$ then for any $A>0$ there are intervals of $\omega$-values arbitrarily close to $0$ for which the dynamics is described by the neutrally stable scenario. The reason for this is as follows:

If there does not exist a $\frac{2\pi}{\omega}$-periodic solution, the system must be in the neutrally stable scenario. By classical Poincar\'{e}-Denjoy theory, the existence or non-existence of $\frac{2\pi}{\omega}$-periodic solutions can be determined by the \emph{asymptotic rotation number}
\[ \Omega \ := \ \lim_{t \to \infty} \frac{\hat{\theta}(t)}{t} \]
where $\hat{\theta} \colon \mathbb{R} \to \mathbb{R}$ is any lift of any solution of \eqref{general}; the value of $\hat{\theta}(0)$ does not affect the value of $\Omega$. There exists a $\frac{2\pi}{\omega}$-periodic trajectory if and only if $\Omega$ is an integer multiple of $\omega$. Now it is well-known that $\Omega$ depends continuously on parameters---in this case, if we fix $k$ and $a$, then $\Omega$ depends continuously on $A$ and $\omega$. But also, observe that $\Omega \in [k-a,k+a]$. Therefore, if $k>a$, then $\frac{\Omega}{\omega}$ must tend continuously towards $\infty$ as $\omega \to 0$ regardless of the value of $A$ (even if $A$ is not actually a fixed value but varies as a continuous function of $\omega$). Hence in particular, there must be intervals of $\omega$-values arbitrarily close to $0$ for which $\frac{\Omega}{\omega}$ is not an integer and so \eqref{general} is in the neutrally stable scenario.

\section*{Numerics for Fig.~1}

The forcing term $g(t)$ is constructed from a \emph{Brownian bridge}. A Brownian bridge effectively describes the result of conditioning a finite-time zero-drift Brownian motion on the event that the start and end values are the same. For the construction of $g(t)$, we start by simulating a realisation of Brownian motion $(W_t)_{0 \leq t \leq T}$, $T=2\pi \times 10^5\;$s, with $W_t \sim \mathcal{N}(0,\frac{t}{T})$. We then construct the Brownian bridge realisation $(B_t)_{0 \leq t \leq T}$ by $B_t := W_t - \frac{t}{T}W_T$. We pass the signal $(B_t)_{0 \leq t \leq T}$ through a 5th order Butterworth lowpass filter with cut-off frequency $1/(2\pi \times 10^3)\;$Hz. The output is $g(t)$.

Numerically, we used a time step of $0.01\;$s to construct the Brownian bridge, and the Butterworth filter was performed via cascaded second-order sections (in Python, with the function ``scipy.signal.sosfilt"). Finally, we linearly interpolated the output of the filter to get $g(t)$.

\begin{figure*}[h!]
	\centering	\includegraphics[width=\linewidth]{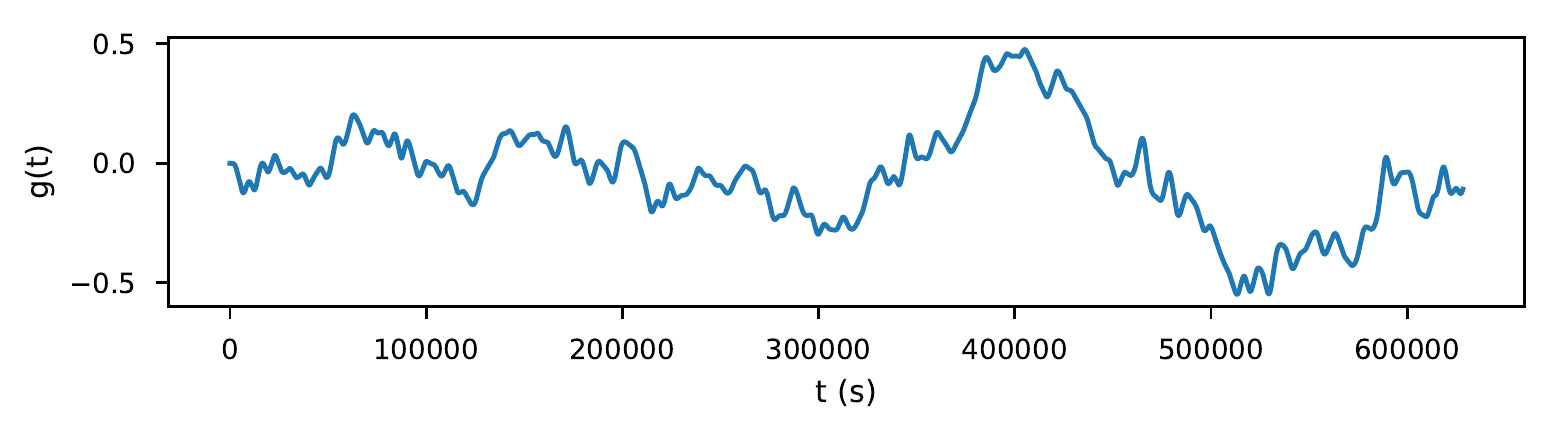}
	\caption{The forcing term $g(t)$ in Eq.~(3).}
	\label{gt}
\end{figure*}

The resulting function $g(t)$ is shown in Fig.~\ref{gt}.

The results shown in Fig.~1(a,b) in the main text were obtained by numerically integrating Eq.~(3) using a 4th order Runge-Kutta scheme, with a time step of $0.01\;$s. The finite-time Lyapunov exponents (FTLE) shown in (a) were calculated according to Eq.~(5). The results in Fig.~1(c) were obtained by evolving the 50 points $\frac{2\pi i}{50}$ under the time-reversed version of (3), namely the differential equation
\[ \dot{\theta}(t) \ = \ a\sin(\theta) - k - Ag(T-t). \]
This was integrated using the same scheme and time step as for (a,b).

\section*{Further numerics of (3)}

\begin{figure*}[h!]
	\centering	\includegraphics[width=\linewidth]{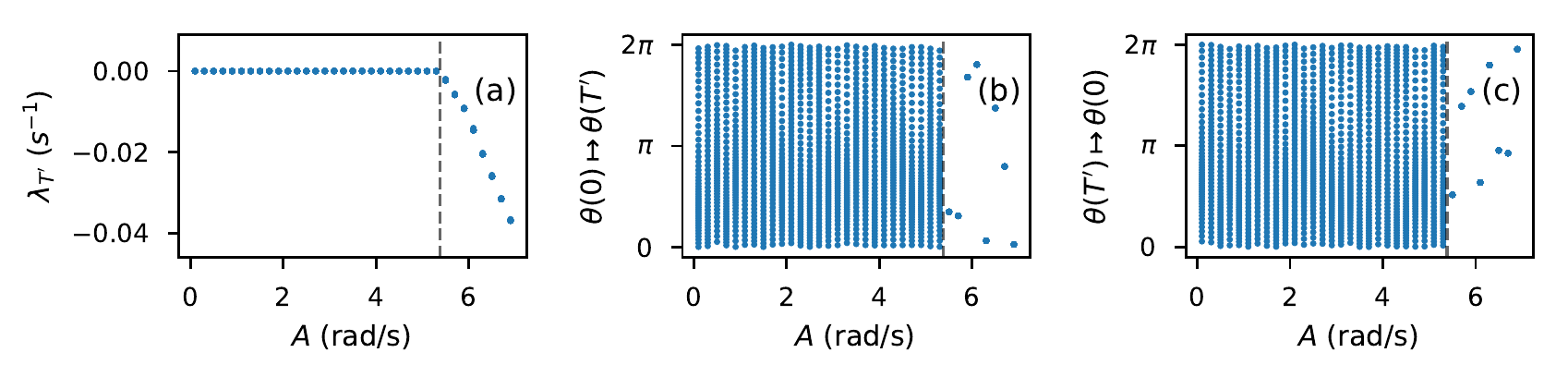}
	\caption{Stabilisation of (3) on time $[0,T']$ with $T'=\pi \times 10^4\;$s. Other parameters are $a=\frac{1}{3}\;\mathrm{rad}/\mathrm{s}$ and $k=1\;\mathrm{rad}/\mathrm{s}$, as in the main text. In (a) and (b), for each $A$-value, results for the evolution $\theta(t)$ of 50 equally spaced initial conditions $\theta(0)=\frac{2\pi i}{50}$, $i=0,\ldots,49$, are shown: (a) shows the FTLE $\lambda_{T'}$, as defined by Eq.~(5), for these trajectories; (b) shows the positions $\theta(T')$ of these trajectories at time $T'$. In (c), for each $A$-value, the positions of $\theta(0)$ for the 50 trajectories of (3) with $\theta(T')=\frac{2\pi i}{50}$, $i=0,\ldots,49$, are shown. The value $A^\ast$ as defined in \eqref{critsub} is marked by the black dashed line.}
	\label{gshort}
\end{figure*}

We now show the same stabilisation phenomenon occurring when we consider (3) not on the whole time-interval $[0,T]$ but just on a subinterval $[0,T']$, where $T'=\pi \times 10^4\;$s. We use the same function $g(t)$, only going up to time $T'$. This function still attains negative values on the subinterval $[0,T']$, and so the critical $A$-value is now given by
\begin{equation} \label{critsub} A^\ast \ = \  \frac{a-k}{\min_{0 \leq t \leq T'} \, g(t)}. \tag{***} \end{equation}
Results are shown in Fig.~\ref{gshort}, with $a$ and $k$ as in Fig.~1 of the main text. Plots~(a) and (b) are obtained exactly as for Fig.~1 in the main text, and likewise plot~(c) by evolving the 50 points under the differential equation
\[ \dot{\theta}(t) \ = \ a\sin(\theta) - k - Ag(T'-t). \]

\section*{Numerics for Fig.~2}

For both the main text and the numerics below, Eq.~(6) was integrated using the 4th order Runge-Kutta scheme with time step $0.01\;$s (just as for (3)), and FTLE were computed according to Eq.~(5). The value of $\tilde{\Lambda}$ marked in grey in Fig.~2(a,b), as defined by Eq.~(7), is given by $\tilde{\Lambda} = -\frac{1}{\pi} \int_{\mathrm{arc}\cos(-\frac{2}{3})}^\pi \sqrt{\frac{1}{9}-(1+\cos(t))^2} \, dt$.

In Fig.~2(b) of the main text, the $\omega$-values at which the red points are marked were numerically obtained as follows: For the unwrapped phase $x(t)$ as governed by the differential equation
\[ \dot{x}(t) \ = \ -a\sin(x(t)) + k + A\cos(\omega t) \]
on the real line, setting $x(0)=0$, it was observed that $x(\frac{2\pi}{\omega})$ increased approximately linearly with $1/\omega$, with increments across consecutive values in the $(1/\omega)$-discretisation being strictly positive and very small compared to $2\pi$. Hence it is possible to carry out linear interpolation of the wrapped phase $\theta(\frac{2\pi}{\omega})$ as a function of $1/\omega$ (with $\theta(0)=0$). Where this linearly interpolated function of $1/\omega$ crosses $\frac{\pi}{2}$ and $\frac{3\pi}{2}$ is where the red points are marked; as in Fig.~2(c), the locations of $\theta(\frac{2\pi}{\omega})$ are the same for the other 49 initial conditions $\theta(0)=\frac{2\pi i}{50}$ as for $\theta(0)=0$.

Let us now explain the reasoning behind why these points indicate the location of Neutrally Stable Scenario intervals:

Let $f_{0,\frac{2\pi}{\omega}} \colon \mathbb{S}^1 \to \mathbb{S}^1$ be the map sending an initial condition $\theta(0)$ to its position $\theta(\frac{2\pi}{\omega})$ at time $\frac{2\pi}{\omega}$. It is not hard to show that the reflection $\theta \mapsto \pi-\theta$, i.e.\ the reflection preserving the points $\frac{\pi}{2}$ and $\frac{3\pi}{2}$, is a conjugacy between $f_{0,\frac{2\pi}{\omega}}$ and its inervse $f_{\frac{2\pi}{\omega},0}$.

By {[43,~Theorems~1 and 4]}, $f_{0,\frac{2\pi}{\omega}}$ either has: (i) no fixed points, corresponding to the Neutrally Stable Scenario; (ii) two fixed points $s_\omega$ and $p_\omega=\pi-s_\omega$, with $s_\omega$ attracting and $p_\omega$ repelling, corresponding to the Stable Scenario; or (iii) exactly one fixed point $p_\omega \in \{\frac{\pi}{2},\frac{3\pi}{2}\}$, corresponding to the Boundary Scenario. Since $A$ lies in the interval $(k-a,k+a)$, Eq.~(6) defined on the time-interval $[0,\frac{2\pi}{\omega}]$ fulfils all the assumptions of Proposition~2 (with $F(\theta,\tau):=-a\sin(\theta)+k+A\cos(\pi\tau)$). Hence, for small enough $\omega$, it is guaranteed that the map $f_{0,\frac{2\pi}{\omega}}$ has nearly zero gradient throughout the circle minus a tiny arc $P_\omega$, and maps $\mathbb{S}^1 \setminus P_\omega$ onto some tiny arc $S_\omega$. Note that the reflection $\pi - S_\omega$ of $S_\omega$ is contained in $P_\omega$, and that if the arcs $S_\omega$ and $P_\omega$ do not overlap then (6) is in the Stable Scenario with $s_\omega \in S_\omega$ and $p_\omega \in P_\omega$. Conversely, whenever (6) is in the Stable Scenario, we have that $p_\omega \in P_\omega$, and therefore $s_\omega \in \pi - P_\omega$.

The locations of $S_\omega$ and $P_\omega$ are represented in Fig.~2(c) of the main text by a hollow circle and a solid circle respectively. As $1/\omega$ increases, $S_\omega$ moves anticlockwise and $P_\omega$ moves clockwise. As these small arcs cross past each other---which is the same as when they cross past $\frac{\pi}{2}$ or $\frac{3\pi}{2}$---there must be a point at which the attracting and repelling fixed points of $f_{0,\frac{2\pi}{\omega}}$ collide. At the moment of collision, the system is in the Boundary Scenario. As $1/\omega$ is increased beyond this point, before the system can return to the Stable Scenario there must be some interval of $(1/\omega)$-values on which $f_{0,\frac{2\pi}{\omega}}$ has no fixed points, corresponding to the Neutrally Stable Scenario.

\section*{Further numerics of (6)}

In all that follows, Eq.~(6) is integrated using the 4th order Runge-Kutta scheme with time step $0.01\;$s; computation of initial positions $\theta(0)$ given final positions $\theta(T)$ (where $T$ is a multiple of $\frac{2\pi}{\omega}$) is achieved by evolving the given values of $\theta(T)$ under the time-reversed system
\[ \dot{\theta}(t) \ = \ a\sin(\theta) - k - A\cos(\omega t) \]
using the same scheme and time step. FTLE are computed according to Eq.~(5). The value of $\tilde{\Lambda}$ in (7) is computed by the explicit formula

\[
\tilde{\Lambda} \ = \ \left\{ \begin{array}{l l}
\ \ 0 & k>a \textrm{ and } A \leq k-a \\
-\frac{1}{\pi} \int_0^\pi \sqrt{a^2 - (k+A\cos(t))^2} \, dt & k<a \textrm{ and } A < k-a \\
-\frac{1}{\pi} \int_{\mathrm{arc}\cos\left(\frac{a-k}{A}\right)}^\pi \sqrt{a^2 - (k+A\cos(t))^2} \, dt & \textit{either } k \geq a \textrm{ and } k-a<A<k+a, \\ & \ \ \textit{or } k<a \textrm{ and } a-k \leq A < a+k \\
-\frac{1}{\pi} \int_{\mathrm{arc}\cos\left(\frac{a-k}{A}\right)}^{\mathrm{arc}\cos\left(-\frac{k+a}{A}\right)} \sqrt{a^2 - (k+A\cos(t))^2} \, dt & A \geq k+a
\end{array} \right.
\]
where $a$, $k$ and $A$ are all assumed to be nonnegative.

\begin{figure*}[h!]
	\centering	\includegraphics[width=\linewidth]{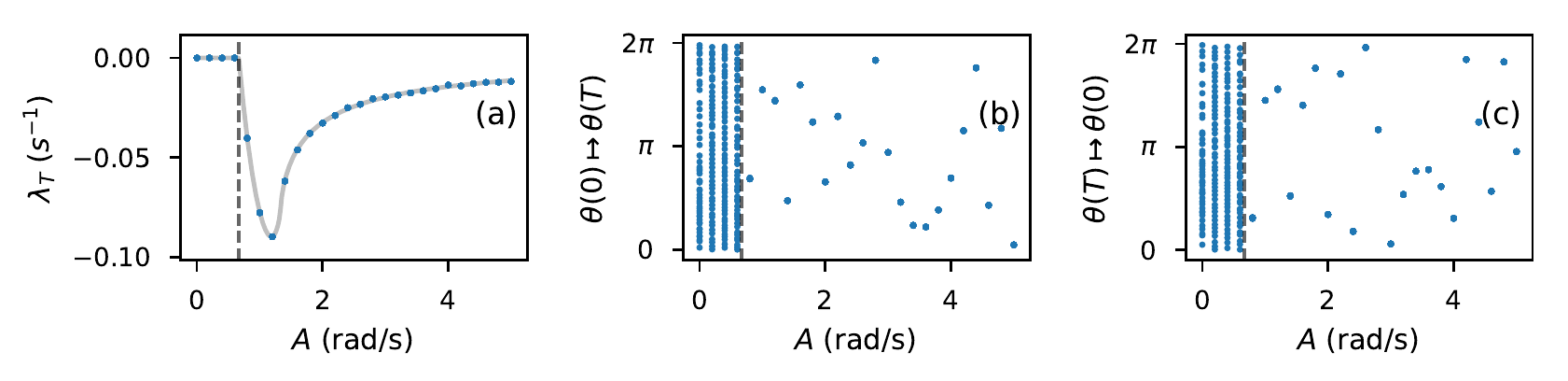}
	\caption{Dynamics of (6) with varying $A$. Other parameters are $\omega = 10^{-3}\;\mathrm{rad}/\mathrm{s}$, $a=\frac{1}{3}\;\mathrm{rad}/\mathrm{s}$ and $k=1\;\mathrm{rad}/\mathrm{s}$, and (6) is integrated over $[0,T]$ with $T=2\pi \times 10^5\;$s. In (a) and (b), for $A$-value, results for the evolution $\theta(t)$ of 50 equally spaced initial conditions $\theta(0)=\frac{2\pi i}{50}$, $i=0,\ldots,49$, are shown: (a) shows the FTLE $\lambda_T$, as defined by Eq.~(5), for these trajectories, and also shows $\tilde{\Lambda}$ in grey; (b) shows the positions $\theta(T)$ of these trajectories at time $T$. In (c), for each $A$-value, the positions of $\theta(0)$ for the 50 trajectories of (6) with $\theta(T)=\frac{2\pi i}{50}$, $i=0,\ldots,49$, are shown. The value $k-a=\frac{2}{3}\;\mathrm{rad}/\mathrm{s}$ is marked by the black dashed line.}
	\label{varA}
\end{figure*}

\begin{figure*}[h!]
	\centering	\includegraphics[width=\linewidth]{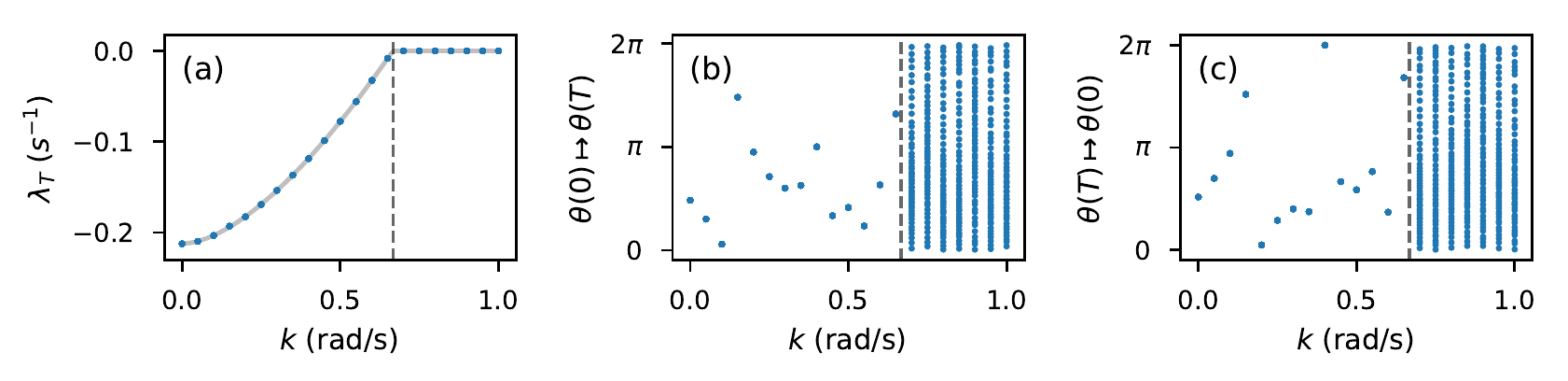}
	\caption{Dynamics of (6) with varying $k$. Other parameters are $\omega = 10^{-3}\;\mathrm{rad}/\mathrm{s}$ and $A=a=\frac{1}{3}\;\mathrm{rad}/\mathrm{s}$, and (6) is integrated over $[0,T]$ with $T=2\pi \times 10^5\;$s. In (a) and (b), for each $k$-value, results for the evolution $\theta(t)$ of 50 equally spaced initial conditions $\theta(0)=\frac{2\pi i}{50}$, $i=0,\ldots,49$, are shown: (a) shows the FTLE $\lambda_T$, as defined by Eq.~(5), for these trajectories, and also shows $\tilde{\Lambda}$ in grey; (b) shows the positions $\theta(T)$ of these trajectories at time $T$. In (c), for each $k$-value, the positions of $\theta(0)$ for the 50 trajectories of (6) with $\theta(T)=\frac{2\pi i}{50}$, $i=0,\ldots,49$, are shown. The value $A+a=\frac{2}{3}\;\mathrm{rad}/\mathrm{s}$ is marked by the black dashed line.}
	\label{vark}
\end{figure*}

Figs.~\ref{varA} and \ref{vark} show the dynamics of (6) for varying $A$ and for varying $k$, respectively. In both cases, $\omega = 10^{-3}\;\mathrm{rad}/\mathrm{s}$ (corresponding to the same frequency as the cut-off frequency in the filter used for constructing $g(t)$), and $T=\frac{200\pi}{\omega}=2\pi \times 10^5\;$s (the same duration as for (3) in Fig.~1).

In Fig.~\ref{varA}, we see neutrally stable behaviour when $A<k-a$ and stable behaviour when $A>k-a$. Exactly the same holds in Fig.~\ref{vark}, where we see neutrally stable behaviour when $k>A+a$ and stable behaviour when $k<A+a$. In both cases, the FTLE are approximated well by $\tilde{\Lambda}$.

One interesting feature in Fig.~\ref{vark}(a) is that the left-sided derivative of $\tilde{\Lambda}$ as a function of $k$ is \emph{finite} at the critical value $k=A+a$ between stable and neutrally stable behaviour. This stands in contrast with classical saddle-node bifurcation at $k=a$ for the autonomous system where $A=0$: in the autonomous case, for each $k<a$ the stable fixed point has Lyapunov exponent $\lambda(k)=-\sqrt{a^2-k^2}$, whose gradient as a function of $k$ grows to \emph{infinity} as $k \nearrow a$.

\begin{figure*}[h!]
	\centering	\includegraphics[width=9cm]{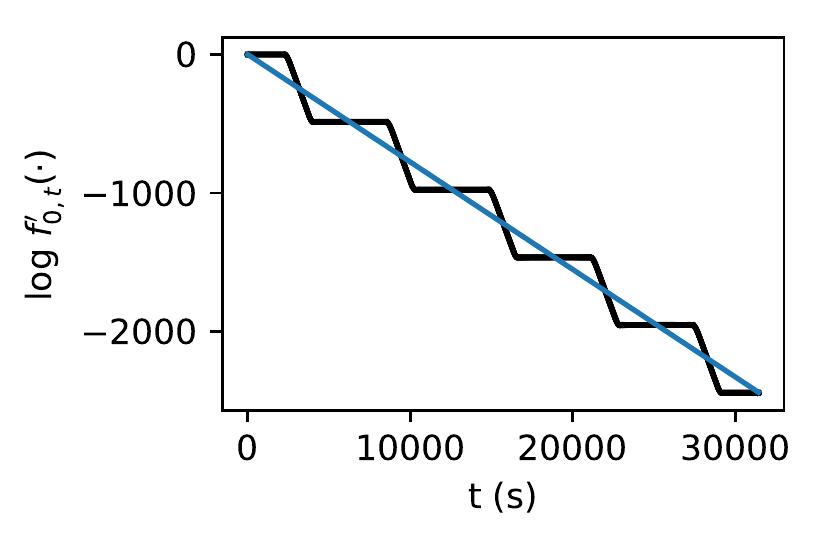}
	\caption{Evolution of synchrony of solutions of (6), over the time-interval $[0,\frac{10\pi}{\omega}]$, i.e.\ 5 time-periods. Parameters are $\omega = 10^{-3}\;\mathrm{rad}/\mathrm{s}$, $k=A=1\;\mathrm{rad}/\mathrm{s}$ and $a=\frac{1}{3}\;\mathrm{rad}/\mathrm{s}$. For each $t$-value, the values of $\log \, f_{0,t}'(\theta_0)$ for 50 equally spaced points $\theta_0=\frac{2\pi i}{50}$, $i=0,\ldots,49$, are shown in black, where $f_{0,t} \colon \mathbb{S}^1 \to \mathbb{S}^1$ is the map sending an initial condition to its position at time $t$. Also, for each $t$-value, the value of $t\tilde{\Lambda}$ is shown in blue.}
	\label{vart}
\end{figure*}

Now, let us illustrate the synchronising dynamics developing over time, for $k=A=1\;\mathrm{rad}/\mathrm{s}$ and $a=\frac{1}{3}\;\mathrm{rad}/\mathrm{s}$, again with $\omega = 10^{-3}\;\mathrm{rad}/\mathrm{s}$. Fig.~\ref{vart} shows behaviour over the first 5 periods of $\cos(\omega t)$: above each $t$-value are shown the values of $\log f_{0,t}'(\theta_0)$ for 50 equally spaced points $\theta_0=\frac{2\pi i}{50}$, $i=0,\ldots,49$, where $f_{0,t} \colon \mathbb{S}^1 \to \mathbb{S}^1$ is the map sending an initial condition to its position at time $t$. These values are computed by
\[ \log \, f_{0,t}'(\theta_0) \ = \ t\lambda_t \]
where, for each $t$, $\lambda_t$ is the FTLE as in Eq.~(5) with $\theta(0)=\theta_0$. In agreement with the description given in the main text, all the trajectories exhibit neutrally stable evolution until some time when they start to synchronise, corresponding to when the instantaneous vector field has a fixed point; the achieved synchrony is maintained during the next time-interval corresponding to when there is no fixed point for the instantaneous vector field; this synchrony is then strengthened further during the next time-interval corresponding to when there is a fixed point again; and so on. Again, we see $\tilde{\Lambda}$ being a good prediction for the FTLE over integer time-periods.

\begin{figure*}[h!]
	\centering	\includegraphics[width=17cm]{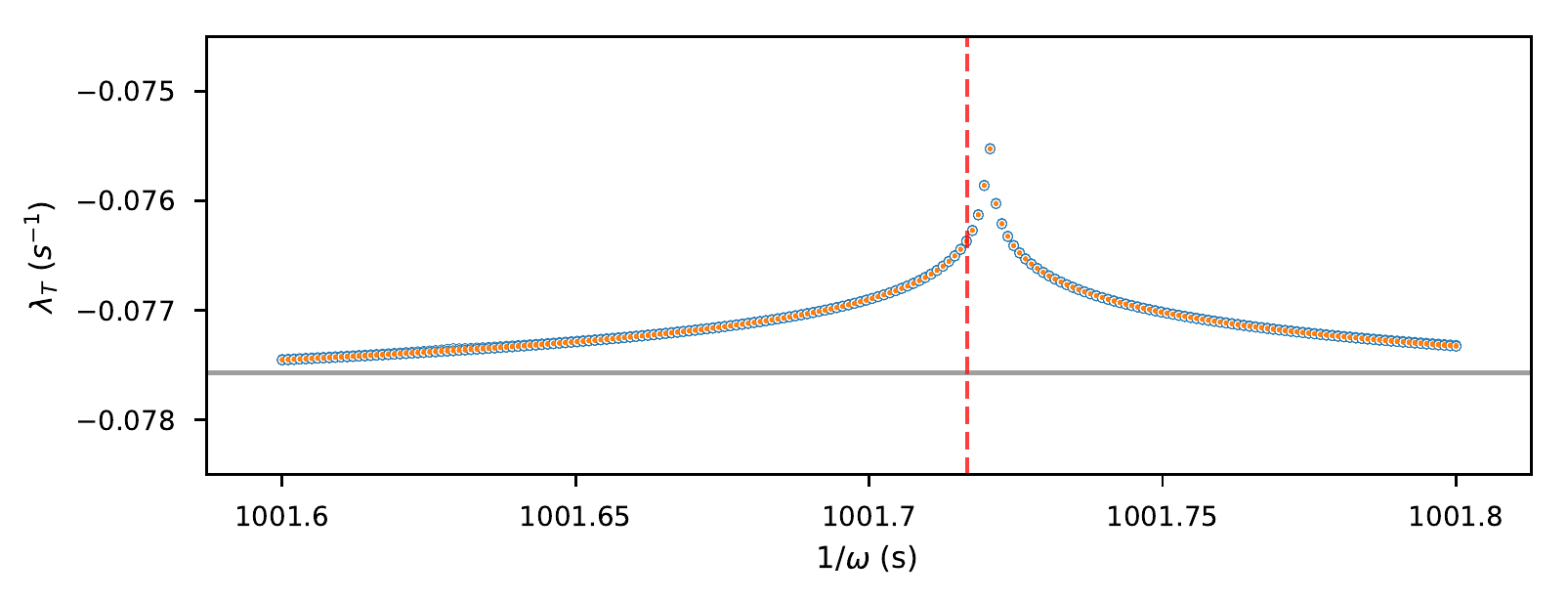}
	\caption{FTLE for (6) near a red-marked point in Fig.~2(b) of the main text. Parameters are $k=A=1\;\mathrm{rad}/\mathrm{s}$ and $a=\frac{1}{3}\;\mathrm{rad}/\mathrm{s}$. For each $\omega$-value, in blue are shown the FTLE $\lambda_T$ associated to the trajectories of 10 initial conditions $\theta(0)=\frac{2\pi i}{10}$ with $T=\frac{200\pi}{\omega}$ (i.e.\ 100 periods), and in orange are shown the FTLE $\lambda_T$ associated to the trajectories of 2 initial conditions $\theta(0)=0,\pi$ with $T=\frac{2000\pi}{\omega}$ (i.e.\ 1000 periods). The red dashed line indicates the location of a small interval of $\omega$-values for which the asymptotic Lyapunov exponent is $0$. The value of $\tilde{\Lambda}$ is shown in grey.}
	\label{weirdpoint}
\end{figure*}

Now Fig.~2(b) of the main text showed FTLE as a function of $\omega$. We now zoom in on Fig.~2(b), near one of the $\omega$-values that was marked by a red point to indicate the presence of a small interval of $\omega$-values for which the asymptotic dynamics is described by the Neutrally Stable Scenario. Results are shown in Fig.~\ref{weirdpoint}. The location of this small interval is indicated by the red dashed line; the location on this zoomed in plot was computed by the same method as described above for Fig.~2(b).

In Fig.~\ref{weirdpoint}, even with the much higher $(1/\omega)$-resolution than in Fig.~2(b), only stable dynamics is observed for both $T=\frac{200\pi}{\omega}$ (as in Fig.~2(b)) and $T=\frac{2000\pi}{\omega}$; moreover, the values of $\lambda_T$ remain essentially the same as $T$ is changed from $\frac{200\pi}{\omega}$ to $\frac{2000\pi}{\omega}$.

\begin{figure*}[h!]
	\centering	\includegraphics[width=\linewidth]{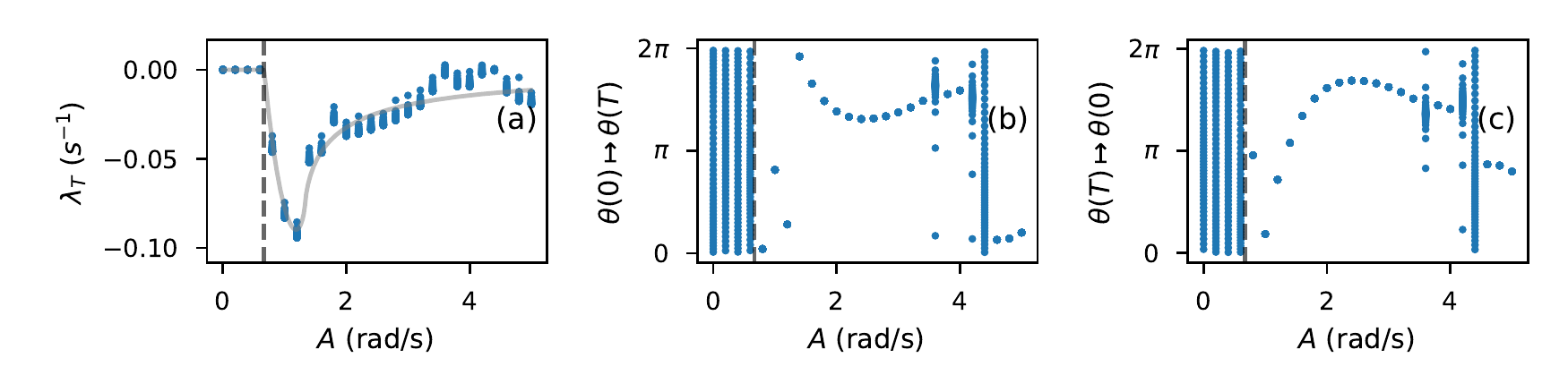}
	\caption{Dynamics of (6) with varying $A$. Other parameters are $\omega = 0.03\;\mathrm{rad}/\mathrm{s}$, $a=\frac{1}{3}\;\mathrm{rad}/\mathrm{s}$ and $k=1\;\mathrm{rad}/\mathrm{s}$, and (6) is integrated over $[0,T]$ with $T=\frac{10\pi}{\omega}$. In (a) and (b), for each $A$-value, results for the evolution $\theta(t)$ of 50 equally spaced initial conditions $\theta(0)=\frac{2\pi i}{50}$, $i=0,\ldots,49$, are shown: (a) shows the FTLE $\lambda_T$, as defined by Eq.~(5), for these trajectories, and also shows $\tilde{\Lambda}$ in grey; (b) shows the positions $\theta(T)$ of these trajectories at time $T$. In (c), for each $A$-value, the positions of $\theta(0)$ for the 50 trajectories of (6) with $\theta(T)=\frac{2\pi i}{50}$, $i=0,\ldots,49$, are shown. The value $k-a=\frac{2}{3}\;\mathrm{rad}/\mathrm{s}$ is marked by the black dashed line.}
	\label{faster}
\end{figure*}

Finally: Most of the numerics so far have assumed very slow variation---which, for (6), means that $A\omega$ is very small---and assumed very long times. We now illustrate that the stabilisation phenomenon described in the main text can be observed when the slowness of variation is not so extreme, and the time is not so long. Fig.~\ref{faster} shows the dynamics of (6) for varying $A$, with $\omega = 0.03\;\mathrm{rad}/\mathrm{s}$ and $T=\frac{10\pi}{\omega} \approx 10^3\;$s (i.e.\ 5 time-periods), again with $a=\frac{1}{3}\;\mathrm{rad}/\mathrm{s}$ and $k=1\;\mathrm{rad}/\mathrm{s}$. We clearly see neutral stability for $A<k-a=\frac{2}{3}\;\mathrm{rad}/\mathrm{s}$. At the point that $A$ rises above $k-a$, we clearly see stabilisation occurring; this stability persists for $A$-values up to about $3.4\;\mathrm{rad}/\mathrm{s}$.

